\documentclass[%
 reprint,longbibliography,preprintnumbers,
nofootinbib,
 amsmath,amssymb,
 aps,
pre,
]{revtex4-1}
\pdfoutput=1
\usepackage{flushend}
\usepackage{dcolumn}
\usepackage{bm}
\usepackage{balance}
\usepackage{tensor}
\usepackage{physics}
\usepackage{wrapfig}
\usepackage{subcaption}
\usepackage{graphicx}

\usepackage[normalem]{ulem}

\usepackage[colorlinks = true,
            linkcolor = blue,
            urlcolor  = blue,
            citecolor =green,
            anchorcolor = blue]{hyperref}
\usepackage{verbatim}
\usepackage{color,ulem}
\usepackage[english]{babel}

\usepackage[utf8]{inputenc}

\newcommand{\beq}{\begin{eqnarray}}
\newcommand{\eeq}{\end{eqnarray}}

\newcommand{\cosOb}[1]{\cos\left(\bar{\omega}_i\left(#1\right)\right)}
\newcommand{\sinOb}[1]{\sin\left(\bar{\omega}_i\left(#1\right)\right)}
\renewcommand{\vec}[1]{\underline{#1}}
\usepackage{amsmath}
\usepackage{tikz}
\usetikzlibrary{decorations.pathmorphing}
\usetikzlibrary{shapes.misc}
\tikzset{cross/.style={cross out, draw=black, minimum size=8*(#1-\pgflinewidth), inner sep=0pt, outer sep=0pt},
cross/.default={1pt}}
\usetikzlibrary{patterns,math}
\begin{document}

\title{Translation-invariant relativistic Langevin equation derived from first principles}

\author{\textbf{Filippo Zadra}$^{1}$}%
\email{filippoemanuele.zadra@studenti.unimi.it}
\author{\textbf{Aleksandr Petrosyan}$^{2}$}%
\email{ap886@cantab.cam.ac.uk}
\author{\textbf{Alessio Zaccone}$^{1,3}$}%
\email{alessio.zaccone@unimi.it}

\vspace{1cm}

\affiliation{$^{1}$Department of Physics ``A. Pontremoli'', University of Milan, via Celoria 16, 20133 Milan, Italy.}
\affiliation{$^{2}$Cavendish Laboratory, University of Cambridge, JJ Thomson Avenue, CB3 0HE, Cambridge, United Kingdom}
\affiliation{$^{3}$ I. Physikalisches Institut, University of G\"ottingen, Friedrich-Hund-Platz 1, 37077 G\"ottingen, Germany}

\begin{abstract}
    The relativistic Langevin equation poses a number of technical and conceptual problems related to its derivation and underlying physical assumptions. Recently, a method has been proposed in [A. Petrosyan and A. Zaccone, J. Phys. A: Math. Theor. 55 015001 (2022)] to derive the relativistic Langevin equation from a first-principles particle-bath Lagrangian. As a result of the particle-bath coupling, a new ``restoring force'' term appeared, which breaks translation symmetry. Here we revisit this problem aiming at deriving a fully translation-invariant relativistic Langevin equation. We successfully do this by adopting the renormalization potential protocol originally suggested by Caldeira and Leggett. The relativistic renormalization potential is derived here and shown to reduce to Caldeira and Leggett's form in the non-relativistic limit. The introduction of this renormalization potential successfully removes the restoring force and a fully translation-invariant relativistic Langevin equation is derived for the first time.
    The physically necessary character of the renormalization potential is discussed in analogy with non-relativistic systems, where it emerges due to the renormalization of the tagged particle dynamics due to its interaction with the bath oscillators (a phenomenon akin to level-repulsion or avoided-crossing in condensed matter).
    We discuss the properties that the corresponding non-Markovian friction kernel has to satisfy, with implications ranging from transport models of the quark-gluon plasma, to relativistic viscous hydrodynamic simulations, and to electrons in graphene. 
\end{abstract}

\maketitle
\section{Introduction}
Brownian motion as described by Langevin equations \cite{langToDiff} has widespread applications from liquid dynamics, to chemical physics, to nuclear physics.
Moreover, these models serve as the basis for the important fluctuation-dissipation theorem (FDT) \cite{zwanzig} and extend to hydrodynamics \cite{langToHydro}.
Thus, it is vital to explore their potential in investigating high-energy fluids, which also encompasses the description of plasmas under appropriate conditions \cite{Zhang}.
However, dealing with high-energy physics presents a challenge as it typically requires the utilization of relativistic frameworks, such as general relativity or special relativity \cite{mullins2023relativistic}. This motivates the need to formulate appropriate relativistic extension of dissipative equations of motion such as the Langevin equation.

For example, there has been recently a lot of interest in implementing stochastic thermal fluctuations in viscous relativistic simulations of heavy-ion collisions \cite{Singh} and in dissipative models for effective field theory \cite{batini}.
However, the corresponding fluctuation-dissipation theorem in these models is invariably assumed to be Markovian, which cannot be given for granted in the absence of a first-principles derivation of the underlying relativistic Langevin dynamics. In the context of the quark-gluon plasma (QGP), the Langevin transport model has proved useful to describe the diffusion of heavy quarks \cite{Teaney_2005,Teaney_2006,He_2023}, but it relies on the non-relativistic Langevin equation.
Finally, a Langevin description accounting for relativistic effects might be important for electrons in graphene, which are both weakly relativistic and posses viscosity \cite{win2023graphene,Levitov}. 

In the context of the relativistic Langevin equation, two main approaches can be identified.
The first approach is more general and based on a relativistic extension of the Ornstein-Uhlenbeck process \cite{debbasch1997}, while the second approach derives the Langevin equation from more specific frameworks, such as relativistic dissipative hydrodynamics \cite{koide} and dissipation models \cite{relBrown1+1}.
In general, the more specialized the assumptions, the more difficult it is  to generalize the final results beyond their intended framework.
In all of these models, however, substantial assumptions are used, in the absence of a first-principles derivation from an underlying Lagrangian or Hamiltonian.

In light of this issue, Ref. \cite{paper} proposes a first-principle derivation from a (Caldeira-Leggett) particle-bath Hamiltonian. During this process, however a new apparent force emerges, which breaks translation symmetry.
In this paper, we consider this issue by introducing a renormalization potential that absorbs the translation-symmetry breaking term, thereby yielding a translation-invariant relativistic Langevin equation.
Before addressing this objective, we briefly review the existing literature on this subject.
After that, we will showcase the methods employed to solve the problem presented by Ref. \cite{paper} and finally, arrive at a manifestly translation-invariant form of the relativistic Langevin equation.

\section{Previous approaches}
\subsection{Relativistic Ornstein-Uhlenbeck Process}
The first generalization of the Langevin process to special relativity comes from Debbasch et al.\cite{debbasch1997}.
In this paper, the authors presented what they called a ``toy model'' where a Langevin equation of motion valid for relativistic systems is postulated to have a covariant form that correctly recovers the Galilean limit at the low speed.
In later works, this theory was expanded upon to a great extent \cite{debbasch1998,debbasch2001,debbasch2001_2}.

The model is based on two main assumptions:
(i) the stochastic process cannot affect the time component of the system and (ii) the dynamics are described by an Ornstein-Uhlenbeck process. 

Furthermore, as stochastic processes are time dependent\footnote{as they are time-indexed random variables $\{X_t\}$} we are left with the choice of a time parameter.
As \cite{dunkel2009} point out, the natural choice falls on the coordinate time of the lab frame, and we shall call it $t$.

Introducing the four-vector formalism and labeling respectively $x=(x^0,\vec{x})$ and $p=(p^0,\vec{p})$ the four-position and the four-momentum of the tagged particle (TP), the friction tensor $\tensor{a}{^i_j}$ and the Wiener process $B^i(t)=(B^1(t),B^2(t),B^3(t))$ one can write the stochastic differential equations (SDEs) for both the dynamic variables:
\begin{align}
	dx^\mu(t)&=p^\mu/p^0dt \label{dx}\\
	dp^i(t)&=F^i(t)dt-\tensor{a}{^i}{_j}p^j dt+B^i(t) \label{OUp}
\end{align}
where we extended Eq.\eqref{dx} to work also with the time component and $\vec{F}(t)$ is a deterministic external force (e.g. Lorentz force) \cite{dunkel2009}.
Furthermore, the second assumption is verified by the fact that an Ornstein-Uhlenbeck process is a mean-retrieving Wiener process, and Eq.\eqref{OUp} reflects just this fact. 
Importantly, the stochastic force within this approach was postulated (without derivation) to be delta-correlated in time, i.e. a Markovian process. This assumption was shown by later approaches (based on a first-principles derivation) such as \cite{paper} to be not realistic because the causality of particle-bath couplings introduces an inherently non-Markovian character to the fluctuation-dissipation theorem, such that the Markovian case cannot, generally, be retrieved.

\subsection{Specialized approaches}

Starting with the work of Dunkel et al. \cite{relBrown1+1}, a (1+1) dimensional dissipation-based approach was proposed.
The main idea behind the derivation is to assume a TP immersed in a particle heat bath (PHB) of vanishing temperature $T \to 0$, which allows them to drop the stochastic term.

By picking a comoving frame of reference (FoR) $\Sigma_*$ such that at a certain time $t$, $v_*(t)=v_*(t_*(t))=0$, then in general, the PHB will have a non-vanishing speed $V_*$ and as such has to be taken into account, giving:
\begin{align}
    \frac{dv_*}{dt_*}(t)&=-\xi(v_*(t)-V_*)=\xi V_*\\
    \frac{dE_*}{dt_*}(t)&=- m \xi v_*(t)(v_*(t)-V_*)=0.
\end{align}
This serves as the basis to write the four-vector Newton equation, by introducing the proper time $d\tau=dt_*/\gamma(v_*)$ the two equations become one:
\begin{equation}
    \frac{dp_*^\mu}{d\tau}=-m \xi(0,v_*(t)-V_*)=-m\tensor{\xi}{^\mu_\nu}(u_*^\nu-U_*^\nu)
\end{equation}
where $\tensor{\xi}{^\mu_\nu}=\mathrm{diag}(0,\xi)$ and $u^\mu$ and $U^\mu$ are both the TP and PHB four-speeds.
The relativistic Langevin equation is then retrieved by reintroducing the (1+1) Wiener process $w^\mu=(0,w)$ giving the sought-after SDE:
\begin{equation}
    dp^\mu(\tau)=-\tensor{\xi}{^\mu_\nu}(p^\nu(\tau)-mU^\nu)d\tau+w^\mu(\tau)
\end{equation}
where the increments are Gaussian distributed in such a way that the velocity increments remain bounded \cite{relBrown1+1}.
\\

Koide et.al. \cite{koide}, instead focus on introducing causality into the Landau-Lifshits theory of relativistic dissipative hydrodynamics.
The idea is similar to achieving the same goal in the non-relativistic case, which is effectively implemented by limiting the interaction speed.
Taking inspiration from Cattaneo's heat equation \cite{cattaneo},  mollifiers \cite{brezis} are employed to impose limits on the interaction.

The derivation follows by noting that in nonequilibrium thermodynamics there is an irreversible current that is proportional to the driving force and Fick’s law expresses the fact that diffusion is induced by spatially inhomogeneous concentrations of particles.
But this current is ``delayed'' due to the finite speed of light, and so the idea is to ``mollify'' the behavior of the time dependence.
Since entropy production, in nonequilibrium thermodynamics, is the sum of the products of different thermodynamic driving forces and irreversible currents, 
these have to be split up due to Curie's (symmetry) principle.
All that is left, at this point, is to modify the driving forces in such a way that makes them comply with the relativistic causality principle.
This is done, as before, by mollifying the driving forces with respect to the proper time.
\\

\section{The first-principles approach}
In light of the aforementioned issues, Ref. \cite{paper} took inspiration from the work of Zwanzig \cite{zwanzig} and implemented a first-principle approach that we shall now present in its most important points.
\subsection{Lagrangian and equations of motion}
Let us start by defining the system at hand.
The system is composed of a massive TP, whose motion will be extended to special relativity, and a PHB made up of relativistic oscillators (RO). The interaction between TP and PHB is bilinear meaning that no two PHB modes can interact and assumes weak interactions between the particle displacement and the oscillators dynamics.
The whole process is adiabatic, in the sense that the TP only exchanges work with the PHB by mechanically interacting with it.
Furthermore, we can take into account an external deterministic force acting on the TP, however, this has to satisfy the causality principle and, as such, will be built on the model of Lorentz force.
And finally, we will choose, as the time parameter, the lab time coordinate $t$ and denote with a dot the lab time derivative.

Having defined the problem at hand, we can start by writing down the Lagrangian for the particle-bath system\footnote{we choose the Lagrangian formalism as the Lagrangian is a pseudo-scalar under Lorentz transformations and Lorentz-covariant Lagrangians yield Lorentz-covariant EOMs \cite{cahill}}. 
For simplicity let us split the Lagrangian up into two components 
\begin{equation}
L=L_{TP}+L_{PHB}
\end{equation}
with
\begin{align}
		L_{\mathrm{TP}}&=\frac{mc^2}{\gamma(\dot{\vec{x}})}-V_{\mathrm{ext}}(\vec{x},\vec{\dot{x}},t)\\
		L_{\mathrm{PHB}}&=\sum_i\frac{m_ic^2}{\gamma(\dot{\vec{q_i}})}-\frac{m_i \omega_i^2}{2\gamma(\vec{\dot{x}})}\norm{q_i-\frac{g_i}{\omega_i^2}x}_4^2 \label{interaction}
\end{align}
where we label $x$ and $q_i$ the position of the TP and of the i-th PHB mode, respectively.
As for the form of the interaction between particle and bath, cfr. the second term on the r.h.s. of Eq. \eqref{interaction}, this is motivated by the following considerations:
\begin{enumerate}
    \item It is the four-vector equivalent of the particle-bath interaction originally used by Zwanzig for classical non-relativistic systems \cite{zwanzig}.
    \item For $c\to\infty$ we correctly retrieve the non-relativistic limit.
    \item It corresponds to the line element $ds^2$ minimizing the variation of the action as it represents the geodesic between the two points in Minkowski space.
\end{enumerate}

Using the Euler-Lagrange equations, we can write the EOMs, for the TP and for the bath oscillators, respectively, as:
\begin{gather}
    \begin{split}
	\frac{d}{dt}\left[\gamma(\dot{\vec{x}})\dot{\vec{x}}\left(m+\sum_i\frac{m_i\omega^2_i}{2c^2} \norm{q_i-\frac{g_i}{\omega_i^2}x}_4^2 \right)\right]=\\
    =-\vec{F}_{ext}+\sum_i\frac{m_ig_i}{\gamma(\dot{\vec{x}})}\left(\vec{q}_i-\frac{g_i}{\omega_i^2}\vec{x}\right) \label{eq:tpEOMnorm}
    \end{split}\\
	\frac{d}{dt}\left[m_i\gamma(\dot{\vec{q}})\dot{\vec{q}}\right]=
	\frac{m_i}{\gamma(\dot{\vec{x}})}\left(\omega_i^2\vec{q}_i-g_i\vec{x}\right). \label{eq:phbEOM}
\end{gather}
We can further simplify Eq.\eqref{eq:tpEOMnorm} by assuming that the interaction is mediated by (light-like) massless particles that, accordingly, follow null-geodesics. This allows us to set $ds^2=0$ and, thus, $\norm{...}_{4}=0$ as the interaction happens instantly within the light cone, leaving us with a more manageable EOM:
\begin{gather}
	\frac{d}{dt}\left[m\gamma(\dot{\vec{x}})\dot{\vec{x}}\right]=-\vec{F}_{ext}+\sum_i\frac{m_ig_i}{\gamma(\dot{\vec{x}})}\left(\vec{q}_i-\frac{g_i}{\omega_i^2}\vec{x}\right) \label{eq:tpEOM}
\end{gather}

Focusing on Eq.\eqref{eq:phbEOM} and expanding the l.h.s. w-e find that the PHB mode trajectories can be decomposed into two subspaces. This goes as follows.

We now can assume to know the trajectory $\vec{x}$ of the TP, and by considering the i-th mode, we notice that $d\gamma({\dot{\vec{q}}})/dt=\gamma^3({\dot{\vec{q}}})\langle\dot{\vec{q}},\ddot{\vec{q}}\rangle_3$.
Defining $\vec{\Lambda}$ as the l.h.s. of Eq. \eqref{eq:phbEOM}, we can expand the time derivative, obtaining:
\begin{align}
		\vec{\Lambda}:=\frac{d}{dt}\left[m_i\gamma(\dot{\vec{q}})\dot{\vec{q}}\right]=\ddot{\vec{q}}_i\gamma(\dot{\vec{q}}_i)+\frac{\gamma^3(\dot{\vec{q}}_i)}{c^2}\langle \dot{\vec{q}}_i,\ddot{\vec{q}}_i \rangle_3.
\end{align}
The newfound EOM for the PHB is now considerably more intricate. However, we can bring it into a simpler form by decomposing the trajectories of the oscillators into longitudinal and transverse components \cite{paper}, where longitudinal here means parallel to the velocity $\dot{\vec{q}}_i$. 
This can be achieved by noting that the particle moves in a linear subspace, we can call this $v_{\parallel}$, and as such $\dot{\vec{q}}_i=\dot{q}_{\parallel_i}\vec{v}_{\parallel_i}$. 
It is good to note that this is an abuse of notation, as $\vec{v}_{\parallel_i}$ represents a vector space and not a formal vector.
We can easily justify this abuse of notation by thinking that the parallel linear subspace has a basis\footnote{where now $v_{\parallel_i}$ does not have the vectorial notation as we are referring to the vector space} $v_{\parallel_i}=\mathrm{span}\{\vec{e}_1,\cdots,\vec{e}_m\}$ and therefore we can compose any vector of said space in terms of the basis vectors by the means of a linear combination, $\vec{w}=w_1\vec{e}_1+\cdots+w_m\vec{e}_m$ where $\norm{\vec{w}}=\dot{q}_i$ and $\vec{w}/\norm{\vec{w}}=\vec{v}_{\parallel_i}$.

We can now decompose any vector into both the parallel subspace $v_{\parallel}$ and the perpendicular subspace $v_{\perp}$ e.g.: $\vec{A}=a\vec{v}_{\parallel_i}+b\vec{v}_{\perp_i}$. Then we can use this decomposition to rewrite the l.h.s. of Eq. \eqref{eq:phbEOM} obtaining:
\begin{align}
		\vec{\Lambda}=(\ddot{q}_{\parallel_i}\vec{v}_{\parallel_i}+\ddot{q}_{\perp_i}\vec{v}_{\perp_i})\gamma(\vec{\dot{q}})+
		\frac{\gamma^3(\vec{\dot{q}})}{c^2}\dot{\vec{q}}^2_{\parallel_i}\ddot{\vec{q}}_{\parallel_i}\vec{v}_{\parallel_i}
\end{align}
By finally projecting\footnote{i.e. by applying the inner product to \eqref{eq:phbEOM}: $\langle\vec{\Lambda},\vec{v}_{\perp_i}\rangle_3=m_i/\gamma(\vec{\dot{x}})\langle(\omega^2_i\vec{\dot{q}}_i-g_i\dot{\vec{x}}),\vec{v}_{\perp_i}\rangle_3$} the EOM onto the two subspaces we get a system of EOMs that describes the dynamics of the PHB oscillators:
\begin{align}
	\gamma^3(\dot{\vec{q}}_i)m_i\ddot{\vec{q}}_{\parallel_i}(t,\vec{x})=
	\frac{m_i}{\gamma(\vec{\dot{x}})}\left(\omega^2_i\vec{q}_{\parallel_i}-g_i\vec{x}_{\parallel_i}\right) 
	\label{eq:EOMparal}\\
	\gamma(\dot{\vec{q}}_i)m_i\ddot{\vec{q}}_{\perp_i}(t,\vec{x})=
	\frac{m_i}{\gamma(\vec{\dot{x}})}\left(\omega^2_i\vec{q}_{\perp_i}-g_i\vec{x}_{\perp}\right).
	\label{eq:EOMperp}
\end{align}

A plane that contains the unperturbed mode motion, called parallel $\vec{q}_\parallel$ and its orthogonal subspace $\vec{q}_\perp$.
So we can rewrite Eq.\eqref{eq:phbEOM} as:
\begin{align}
  \begin{split}
		\frac{d}{dt}\left[m_i\gamma(\dot{\vec{q}})\dot{\vec{q}}\right]&=(\ddot{q}_{\parallel_i}\vec{v}_{\parallel_i}+\ddot{q}_{\perp_i}\vec{v}_{\perp_i})\gamma(\vec{\dot{q}})+\\
		&+\frac{\gamma^3(\vec{\dot{q}})}{c^2}\dot{\vec{q}}^2_{\parallel_i}\ddot{\vec{q}}_{\parallel_i}\vec{v}_{\parallel_i}=
        \frac{m_i}{\gamma(\dot{\vec{x}})}\left(\omega_i^2\vec{q}_i-g_i\vec{x}\right)
  \end{split}
\end{align}
By projecting the equation onto both subspaces (e.g. $\langle \vec{\Lambda}, \hat{q}_\parallel \rangle_3$) we can decompose the trajectory into two EOMs:
\begin{align}
	\gamma^3(\dot{\vec{q}}_i)m_i\ddot{\vec{q}}_{\parallel_i}(t,\vec{x})=
	\frac{m_i}{\gamma(\vec{\dot{x}})}\left(g_i\vec{x}_{\parallel_i}-\omega^2_i\vec{q}_{\parallel_i}\right) 
	\label{eq:EOMparal}\\
	\gamma(\dot{\vec{q}}_i)m_i\ddot{\vec{q}}_{\perp_i}(t,\vec{x})=
	\frac{m_i}{\gamma(\vec{\dot{x}})}\left(g_i\vec{x}_{\perp}-\omega^2_i\vec{q}_{\perp_i}\right).
	\label{eq:EOMperp}
\end{align}
As these two equations have a lot in common, we note that, by Taylor expanding around $\vec{\dot{q}}\to 0$, they share the same solution, albeit with different numerical values, as the Taylor expansions differ only in the numerical coefficients.
Because of this, from now we will now drop the subscripts referring to the orthogonal subspaces.

\subsection{Kernel for the relativistic oscillators bath dynamics}
In the form found above, neither Eq.\eqref{eq:EOMparal} nor Eq.\eqref{eq:EOMperp} can be solved analytically, due to the strong anharmonic nonlinear character of the relativistic oscillators which form the PHB.
Because of this, Ref. \cite{paper} proposed to analyze the numerical solution of the aforementioned equations and compare them with the non-relativistic analytical solution found in \cite{zwanzig}.
By introducing a frequency shift for the bath oscillators to fit the numerical data, it was possible to retain all the usability of the analytical non-relativistic solution \cite{zwanzig} supplemented with the relativistic modifications.
This section will be presented in a (1+1) formalism as the (3+1) generalization follows ``mutatis mutandis''.
\begin{figure}
	\centering
	\begin{subfigure}[b]{.40\textwidth}
		\centering
		\includegraphics[width=\textwidth]{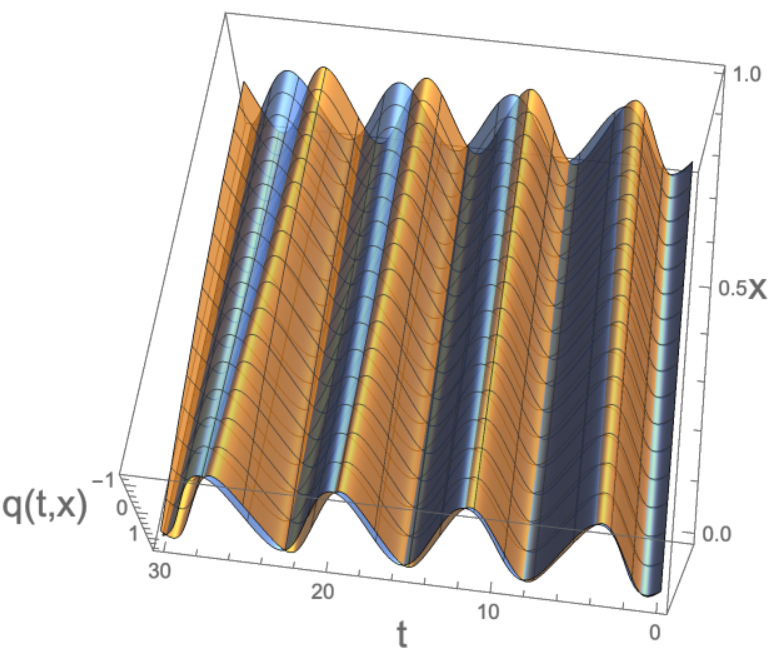}
		\caption{Comparison of the numerical solution (orange) with the non-relativistic solution (blue) as the frequency and phase are set as constant. Adapted from Ref. \cite{paper} with permission of the Institute of Physics Publishing.}
		\label{fig:xshift}
	\end{subfigure}
	\begin{subfigure}[b]{.40\textwidth}
		\centering
		\includegraphics[width=\textwidth]{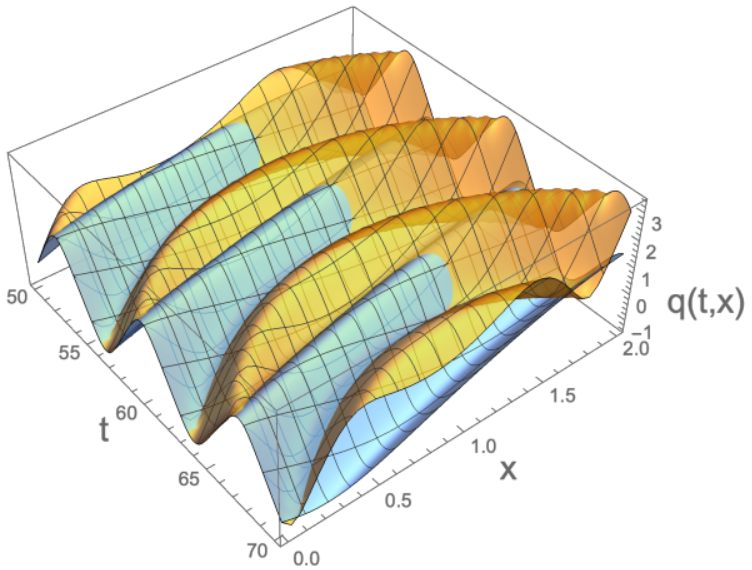}
		\caption{A plot showing the variation from the numerical solution (orange) and the non-relativistic solution (blue) and highlighting the behavior of the system far from the origin}
		\label{fig:wrinkles}
	\end{subfigure}
\end{figure}

In particular, three main effects are evident in Fig.\ref{fig:xshift} and Fig.\ref{fig:wrinkles} and those are:
\begin{enumerate}
		\item{A systematic shift in the eigenfrequency of the bath oscillators as they are ROs, cfr. Fig.\ref{fig:xshift}, which is due to the intrinsic anharmonicity of the ROs;}
		\item{A dependence of the ROs frequency on the TP position $x$ as seen in both Fig.\ref{fig:xshift} and Fig.\ref{fig:wrinkles}}
		\item{Sub-oscillations (``wrinkles'') at the crests of each wave: this phenomenon contributes little to the fitting of the simulations data, hence it will be ignored (see Fig.\ref{fig:wrinkles})}
\end{enumerate}

As explained in Ref. \cite{lifshitz} anharmonic oscillators can be thought of as harmonic oscillator with harmonic eigenfrequency $\omega_i$ plus a correction term that stems from the nonlinear nature of the dynamics of the system. 
In our case, the harmonic eigenfrequency is the one of the classical oscillators with $\gamma(\vec{\dot{q}})=1$ and by comparison with the numerical solution it is found that the correction\footnote{as described before one expects the nonlinear term to be dependent on $\gamma(\dot{\vec{q}})$ since this is what introduced the non-linearity in the first place.} $\delta \omega_i\propto \gamma^{-3/4}(\dot{\vec{q}})$, obtaining a renormalized frequency $\bar{\omega_i}=\omega_i+\delta \omega _i$.

So one proceeds by introducing a frequency shift (or frequency renormalization) $\bar\omega_i=\omega_i+\delta\omega_i(\gamma(\dot{q}))$, and phase-correction coefficient $\xi(\vec{x}(t))$, to take into account the strongly anharmonic, non-linear behavior\footnote{changing effective mass based on speed \cite{teo_RHO}} of the ROs. 

This is in analogy with the self-consistent theory of phonons in condensed matter physics, where the anharmonicity of the interatomic interaction leads to deviations from the harmonic behaviour of the normal modes of vibrations (i.e. the phonons). Also in that case, the deviation from harmonic modes is effectively encoded via a renormalization of the harmonic frequency, exactly in the same way as here \cite{Bao,Casella}.

After that, we can see in Fig.\ref{fig:wrinkles} that the two solutions match at $x=0$. However, now the spatial dependency becomes non-trivial.
In Ref. \cite{paper} the authors took care of this fact, by introducing the relativistic phase-drift correction for the RO oscillations:
\begin{align}
    \bar\xi(t,x)=At(x-B) \label{eq:xi}
\end{align}
where $A, B$ are both parameters that have to be fitted to the numerical solutions for the RO dynamics. As explained in Ref. \cite{paper}, this correction function allows us to absorb the further dependencies on space and time coordinates induced by the RO dynamics into the phase drift correction such that\footnote{where for consistency we use $\bar\omega$ as $\bar\omega\neq\omega$ since the change in eigenfrequencies} $\bar\omega_i=const$.

All of the arguments above lead to the corrected relativistic solution for the RO dynamics:
\begin{equation}
		\begin{split}
			\vec{q}_i(t)&= \vec{q}_i(0)\cosOb{t-\frac{\bar{\xi}_i(t,\vec{x}(t))}{c}}+\\
			&+\frac{1}{\bar{\omega}_i}\gamma(\vec{\dot{q}}_i(0))\vec{\dot{q}}_i(0)\sinOb{t-\frac{\bar{\xi}_i(t,\vec{x}(t))}{c}}+\\
			&+\frac{g_i}{\bar{\omega}_i}\int_0^t \gamma(\vec{\dot{x}}(s))\vec{x}(s) \sinOb{t-\frac{\bar{\xi}_i(t,\vec{x}(t))}{c}-s}ds
			\label{eq:newsol}
		\end{split}
\end{equation}
where we can see the presence of the phase drift correction $\bar\xi$ and the renormalized eigenfrequencies $\bar\omega$.

\subsection{Solving for the relativistic Langevin dynamics: symmetry breaking}
To obtain the solution to the PHB EOMs we have to compute the integral by parts in the last term of Eq.\eqref{eq:newsol}.
Following the same manipulations as in Ref. \cite{paper}, the final form of the solution to Eq.\eqref{eq:phbEOM} is:
\begin{equation}
		\begin{split}
			\vec{q}_i(t)&= \vec{q}_i(0)\cosOb{t-\frac{\bar{\xi}_i(t,\vec{x}(t))}{c}}+\\
			&+\frac{1}{\bar{\omega}_i}\gamma(\vec{\dot{q}}_i(0))\vec{\dot{q}}_i(0)\sinOb{t-\frac{\bar{\xi}_i(t,\vec{x}(t))}{c}}+\\
			&+\frac{g_i}{\bar\omega_i^2}\left[\vec x(t)\int_0^t \gamma(\dot{\vec{x}}(s))\sinOb{t-s-\frac{\bar\xi(s)}{c}}ds\right]_0^t-\\
			&-\frac{g_i}{\bar\omega_i^2}\int_0^t \dot{\vec{x}}(s) \int_0^t \gamma(\dot{\vec{x}}(y))\sinOb{t-s-\frac{\bar\xi(s)}{c}}dyds
			\label{eq:phbEOMsol}
		\end{split}
\end{equation}
In the integral done by part, we have also followed the suggestion of Ref. \cite{paper}, and effectively absorbed the dependence of $\bar{\xi}$ on the trajectory $\vec{x}(s)$ into the (generic) dependence of $\bar{\xi}$ on $s$, since $\vec{x}$ is a parametric function of $s$. Hence, instead of $\bar{\xi}(\vec{x}(s))$ we simply write $\bar{\xi}(s)$.
After some algebraic reshuffling of various terms, we can substitute the solution into Eq.\eqref{eq:tpEOM} and single out the terms which form the (generalized) Langevin equation.
The final result is given by Eq. (30) in Ref. \cite{paper}.
Importantly, the friction kernel satisfies a non-Markovian FDT, which reads as: $\langle \vec{F}'_p(t) \vec{F}'_p(t') \rangle = m k_B T K(t-t')$, where $\vec{F}'_p(t)$ is the stochastic force, $k_{B}$ is the Boltzmann constant and $K(t-t')$ is the friction kernel, and $\langle ... \rangle$ denotes ensemble averaging. As discussed in Ref. \cite{paper}, due to the unavoidable presence of the phase shift $\xi$ and its form discussed above, the friction kernel $K(t-t')$ cannot be reduced to a Dirac delta, which makes the relativistic Langevin equation intrinsically non-Markovian.

Despite successfully deriving a relativistic Langevin equation from a particle-bath first-principle derivation, proving both that the result satisfies a non-Markovian FDT and possesses a manifestly covariant form, an additional symmetry-breaking term arises that Ref. \cite{paper} refers to as the ``restoring force'':

\begin{equation}
    \begin{split}
            \vec{F}_r(t)&=\sum_i\frac{m_ig_i^2\vec{x}(t)}{\gamma(\vec{\dot{x}}(t))\bar\omega_i^2}\times\\
            \times&\biggr[ \int \gamma(\vec{\dot{x}}(s))\bar{\omega}_i \sinOb{t-s-\frac{\bar\xi(s)}{c}}ds - 1 \biggr]_t,
            \label{eq:relrest}
    \end{split}
\end{equation}
where the subscript $t$ indicates that the indefinite integral is evaluated at $s=t$.
However, not only this term should not be present in the standard form of a Langevin equation, but also manifestly breaks translation invariance.
Furthermore, there is no apparent physical motivation for the presence of this term and, even more interestingly, for $c\to \infty$ we have $\vec{F}'_r\to \vec{0}$ further highlighting the hypothesis that this is, indeed, a relativistic effect.

\section{Discussion of the translation-symmetry breaking force}
The objective of this paper is to bring the above relativistic Langevin equation to a manifestly translation invariant form.
To this aim, we first need to dissect the particle-bath Lagrangian used by Zwanzig \cite{zwanzig} and later extended to relativistic settings by Petrosyan and Zaccone \cite{paper}. 
We notice that, as suggested early on by Caldeira and Leggett themselves \cite{CLdiss}, a translation-symmetry breaking term in the dissipative dynamics may arise, under certain conditions, from the particle-bath coupling implemented in the starting Caldeira-Leggett (particle-bath) Lagrangian.

\subsection{Caldeira-Leggett model and the multi-mode Brownian oscillator model}
We start by considering the form of the non-relativistic Lagrangian as it was used by Zwanzig in his original derivation \cite{zwanzig}.
As pointed out by Gottwald et al. \cite{anharmonic}, the Caldeira-Leggett (CL) particle-bath model does not have a renormalized potential, instead the so-called multi-mode Brownian oscillator (MBO) does.

More in detail, a CL heat bath is a bath composed of harmonic oscillators (unable to interact with one another), and of a TP coupled with every oscillator with a linear interaction in position \cite{CLdiss}.
This implies that the total potential energy of the TP should have the following form:
\begin{align}
	V_{\mathrm{CL}}(\vec{x},\{\vec{q}_i\})=V_\mathrm{TP}(\vec{x})+\sum_i\frac{\omega_i^2}{2}\vec{q}_i^2-\sum_ig_i\langle \vec{q}_i,\vec{x} \rangle_3.\label{CL1}
\end{align}

One can then complete the square, and have a total potential energy of the following form:
\begin{align}
	V_{\mathrm{CL}}(\vec{x},\{\vec{q}_i\})=\tilde{V}_\mathrm{TP}(\vec{x})+\sum_i\frac{\omega_i^2}{2}\left( \vec{q}_i - \frac{g_i}{\omega^2_i}\vec{x}\right)^2.\label{CL2}
\end{align}
In this formulation, the new TP potential $\tilde{V}_\mathrm{TP}$ is the renormalized TP potential 
\begin{align}
	\tilde{V}_\mathrm{TP}(\vec{x})\equiv V_\mathrm{TP}(\vec{x}) - \sum_i \frac{g_i^2}{2\omega_i^2}\vec{x}^2.\label{renorm}
\end{align}
Obviously, the two potential energies \eqref{CL1} and \eqref{CL2}, are identical.

If one removes the renormalization of the TP potential, we then have the following model:
\begin{align}
	V_{\mathrm{MBO}}(\vec{x},\{\vec{q}_i\})=V_\mathrm{TP}(\vec{x})+\sum_i\frac{\omega_i^2}{2}\left( \vec{q}_i - \frac{g_i}{\omega^2_i}\vec{x}\right)^2.\label{MBO}
\end{align}
which is the one used by Zwanzig \cite{zwanzig} to the derive the classical generalized Langevin equation, and often referred to as the multi-mode Brownian oscillator (MBO) model.

Both models, i.e. CL and MBO reported above, can be used to derive classical and quantum Langevin equations \cite{anharmonic}.

The need for the use of a renormalized potential $\tilde{V}_\mathrm{TP}(\vec{x})$, and its relation to translation invariance, are discussed in different places in literature, e.g. in \cite{petruccione2005,CLdiss}.
The physical motivation for such a correction term lies in the fact that whenever interaction with the TP happens, a new degree of anharmonicity is introduced into each harmonic oscillator, and as such we have to take into account how the i-th mode trajectory is affected by the interaction.
This effect is well known in quantum mechanics as level-repulsion \cite{CLdiss}, and has important consequences also in condensed matter physics, where a similar coupling between harmonic oscillators and an elastic matrix leads to avoided-crossing (level repulsion) of phonon modes \cite{thermoelectric}.

Caldeira and Leggett go as far as suggesting, in \cite{CLdiss}, that the emergence of a restoring force, such as the one given by Eq. \eqref{eq:relrest} found in Ref. \cite{paper}, is a side effect of the frequency renormalization introduced by the bilinear coupling implemented in the particle-bath Lagrangian.

\subsection{Caldeira-Leggett renormalization protocol}
Following Ref. \cite{CLdiss}, in the following we design a renormalization potential to absorb the translation-symmetry breaking ``restoring force''.
We decided to follow Caldeira and Leggett \cite{CLdiss} as they provide some insights into how to do this, notably in Section 3, where they propose the following protocol:
\begin{enumerate}
	\item Start off with the CL model Eq. \eqref{CL1} in the form where no renormalization potential is present.
	\item Solve for the EOMs and derive the new Langevin equation with the restoring force present.
	\item Compute an ad-hoc potential that cancels out the restoring force.
	\item Add it to the Lagrangian and solve the new EOMs.
	\item If during the derivation process unwanted terms arise, tweak the potential and repeat.
\end{enumerate}

Let us start with the first point, i.e. re-deriving the relativistic Langevin equation using the CL model with the non-renormalized potential.
Accordingly, we write the system Lagrangian:
\begin{gather}
		L_{\mathrm{TP}}=\frac{mc^2}{\gamma(\dot{\vec{x}})}-V_{\mathrm{ext}}(\vec{x},\vec{\dot{x}},t)\label{eq:LTPnonrenorm}\\
		L_{\mathrm{PHB}}=\sum_i\frac{m_ic^2}{\gamma(\dot{\vec{q_i}})}-\frac{m_i \bar\omega_i^2}{2\gamma(\vec{\dot{x}})}\left( \norm{q_i}_4^2 -2\frac{g_i}{\bar\omega^2_i}\langle q_i,x \rangle_4 \right)\label{eq:LPHBnonrenorm}
\end{gather}
Here we decided to keep the four-vector form of the interaction as the presence of the time components does not interfere in the Euler-Lagrange equations. Furthermore, by using a rescaling of the dynamic variable, the interaction potential is indeed manifestly translation-invariant.

As for the EOMs, we have:
\begin{gather}
    \begin{split}
	\frac{d}{dt}\Biggr[\gamma(\dot{\vec{x}})\dot{\vec{x}}\left(m+\sum_i\frac{m_i\bar\omega^2_i}{2c^2}\left( \norm{\vec{q}_i}_3^2 -2\frac{g_i}{\bar\omega^2_i}\langle \vec{q}_i,\vec{x} \rangle_3 \right)\right)\Biggr]&=\\
    =-\vec{F}_{ext}+\sum_i\frac{m_ig_i}{\gamma(\dot{\vec{x}})}\vec{q}_i& \label{eq:tpEOM2}
    \end{split}\\
	\frac{d}{dt}\left[m_i\gamma(\dot{\vec{q}}_i)\dot{\vec{q}}_i\right]=
	\frac{m_i}{\gamma(\dot{\vec{x}})}\left(\bar\omega_i^2\vec{q}_i-g_i\vec{x}\right) \label{eq:phbEOM2}
\end{gather}
We can in fact see that Eq.\eqref{eq:phbEOM2} did not change, as such Eq.\eqref{eq:newsol} is still the right solution for the PHB modes.
Focussing on the l.h.s. of Eq.\eqref{eq:tpEOM2} and using the identity $\norm{a-b}_n^2=\norm{a}_n^2+\norm{b}_n^2-2\langle a,b \rangle_n$ we can cancel the term proportional to the interaction interval\footnote{as we discussed earlier, this is the hypothesis of null line element \cite{paper}} leaving us with:
\begin{align}
	\nonumber
	&\frac{d}{dt}\Biggr[m\gamma(\dot{\vec{x}})\dot{\vec{x}}\left(1+\sum_i\frac{m_i\bar\omega^2_i}{2mc^2}\left( \norm{\vec{q}_i}_3^2 -2\frac{g_i}{\bar\omega^2_i}\langle \vec{q}_i,\vec{x} \rangle_3 \right)\right)\Biggr]=\\
	\nonumber
	=&\frac{d}{dt}\Biggr[m\gamma(\dot{\vec{x}})\dot{\vec{x}}\left(1+\sum_i\frac{m_i\bar\omega^2_i}{2mc^2}\left( \norm{\vec{q}_i-\frac{g_i}{\bar\omega_i^2}\vec{x}}_3^2 - \frac{g_i}{\bar\omega_i^2}\norm{\vec{x}}_3^2 \right)\right)\Biggr]=\\
	=&\frac{d}{dt}\Biggr[m\gamma(\dot{\vec{x}})\dot{\vec{x}}\left(1-\sum_i\frac{m_ig_i}{2mc^2}\vec{x}^2. \right)\Biggr]\label{eq:lhsEOM}
\end{align}

We can now see how we can drop the second term too in the parenthesis of Eq.\eqref{eq:lhsEOM}: this is because of the two hypotheses, namely (i) $m>m_i$ and (ii) by definition of weakly interacting TP with the PHB (as prescribed by the CL model), $g_i\approx 10^0$. These assumptions are certainly valid for:
\begin{align}
	\norm{\vec{x}}_3 \ll \sqrt{\frac{2}{g_i}}c.
	\label{eq:approx}
\end{align}
Then we have the two EOMs:
\begin{gather}
	\frac{d}{dt}\left[m\gamma(\dot{\vec{x}})\dot{\vec{x}}\right]=-\vec{F}_{ext}+\sum_i\frac{m_ig_i}{\gamma(\dot{\vec{x}})}\vec{q}_i \label{eq:tpEOMnorenorm}\\
	\frac{d}{dt}\left[m_i\gamma(\dot{\vec{q}}_i)\dot{\vec{q}}_i\right]=\frac{m_i}{\gamma(\dot{\vec{x}})}\left(\bar\omega_i^2\vec{q}_i-g_i\vec{x}\right) \label{eq:phbEOMnorenorm}
\end{gather}
to make the derivation easier to follow, we can rewrite Eq.\eqref{eq:phbEOMsol} in terms of the i-th contribution to the memory kernel, restoring force and noise function defined in \cite{paper}.
To do this, let us define:
\begin{gather}
	\begin{split}
	\vec{F}_{p_i}=\frac{m_ig_i}{\gamma(\vec{\dot{x}}(t))}&\Biggr( 
		\vec{q}_i(0)\cosOb{t-\frac{\bar{\xi}_i(t)}{c}}-\frac{g_i}{\bar\omega_i}\vec{x}(0)I(0)+\\
		&+\frac{\vec{\dot{q}}_i(0)\gamma(\vec{\dot{q}}_i(0))}{\bar\omega_i}\sinOb{t-\frac{\bar{\xi}_i(t)}{c}}
	\Biggr)
	\end{split}\\
	K_i(t,s)=\frac{g_i}{\bar\omega_i^2\gamma(\vec{\dot{x}}(s))}\int \gamma(\vec{\dot{x}}(y))\sinOb{t-y-\frac{\bar{\xi}_i(t)}{c}}dy\\
	\vec{F}_{r_i}=\frac{g^2_im_i}{\bar\omega_i\gamma(\vec{\dot{x}}(t))}\vec{x}(t)I(t)
\end{gather}
where, to clean up the notation, we defined:
\begin{equation}
	I(t)=\left[\int \gamma(\vec{\dot{x}}(s)) \sinOb{t-s-\frac{\bar{\xi}_i(t)}{c}} ds\right]_t.
	\label{eq:I}
\end{equation}
Then Eq.\eqref{eq:phbEOMsol} can be rewritten as:
\begin{align}
	\frac{m_ig_i}{\gamma(\vec{\dot{x}}(t))}\vec{q}_i(t)=\vec{F}_{p_i}+\vec{F}_{r_i}+\int \vec{x}(t-s)\gamma(\vec{\dot{x}}(t-s))K_i(t,s)ds
	\label{eq:qRescaled}
\end{align}
In this form, it is easy to see how the Langevin equation appears by substituting Eq.\eqref{eq:qRescaled} into Eq.\eqref{eq:tpEOMnorenorm}.
Most importantly, we notice that the restoring force indeed changed, losing the second term in Eq.\eqref{eq:relrest}, which further highlights how the renormalization of the potential indeed affects the restoring force.

\subsection{The relativistic renormalization potential}

Having dealt with points one and two in the Caldeira-Leggett renormalization protocol, in this section, we will finally deduce the form of the renormalization potential and verify that it indeed leads to the vanishing of the restoring force.
Ref. \cite{CLdiss} discusses how this potential has to be dependent on the TP dynamic variables only. If this was not the case, then it would alter the PHB dynamics and EOM, and as such Eq.\eqref{eq:phbEOMsol} would no longer be the solution for the PHB dynamics.

The most natural candidate is:
\begin{align}
	-\vec{\nabla}\Phi(\vec{x},\vec{\dot{x}},t)=\vec{F}_r(\vec{x},\vec{\dot{x}},t)
\end{align}
since, this way, when added to the Lagrangian Eqs. \eqref{eq:LTPnonrenorm},\eqref{eq:LPHBnonrenorm}, the restoring force cancels out, hence the positive sign in the r.h.s.

Then the renormalization potential has the form:
\begin{align}
	\Phi(\vec{x},\vec{\dot{x}},t)=\frac{I(t)}{\gamma(\vec{\dot{x}}(t))} \frac{m_i g_i^{2}}{\bar{\omega}_{i}^{2}} \vec{x}^2(t)
	\label{eq:renormPot}
\end{align}
which is consistent, ``mutatis mutandis'', with its non-relativistic analogue, given by the second term on the r.h.s. in Eq. \eqref{renorm}. In particular, both are quadratic in the position of the TP, and proportional to the factor $\frac{m_i g_i^{2}}{\bar{\omega}_{i}^{2}}$. The only difference is in the (new) relativistic factor $\frac{I(t)}{\gamma(\vec{\dot{x}}(t))}$, which reduces to one in the non-relativistic limit, $\gamma \rightarrow 1$, $c \rightarrow \infty$.

Now using Eqs.\eqref{eq:LPHBnonrenorm},\eqref{eq:LTPnonrenorm},\eqref{eq:renormPot}, we can define $L=L_{\mathrm{TP}}+L_{\mathrm{PHB}}-\Phi(\vec{x},\vec{\dot{x}},t)$ and derive the EOM of the TP, as $\Phi(\vec{x},\vec{\dot{x}},t)$ does not depend on $\vec{\dot{q}}_i$ nor on $\vec{q}_i$ and as such remains unchanged from Eq.\eqref{eq:phbEOM2}.

However, the TP EOM becomes:
\begin{gather}
	\begin{split}
	\frac{d}{dt}&\Biggr[m\gamma(\dot{\vec{x}})\dot{\vec{x}}\left(1-\sum_i\frac{m_ig_i}{2mc^2}\vec{x}^2(1+I(t)) \right)\Biggr]=\\
	&=-\vec{F}_\mathrm{ext}-\vec{F}_r-\sum_i\frac{m_ig_i}{\gamma(\dot{\vec{x}})}\vec{q}_i
	\end{split} \label{newterm}
\end{gather}
where we followed the same steps as in Eq.\eqref{eq:lhsEOM}. 
As we can see upon substituting in Eq. \eqref{eq:qRescaled}, the renormalization potential successfully cancels out the restoring force, however, the l.h.s. acquires a new term in the process, which is given by $\sum_i\frac{m_ig_i}{2mc^2}\vec{x}^2(1+I(t))$ inside the round bracket in the l.h.s. of Eq. \eqref{newterm}. 
We shall now demonstrate that this term vanishes. 
We will now present a qualitative argument, while in the Appendix \ref{app:approx} we will explore a physical justification and analyze the implication of this result.

Based on the assumptions used before in the context of Eq.\eqref{eq:approx}, we have that:
\begin{align}
	\frac{m_ig_i}{2mc^2}\vec{x}^2(1+I(t))\approx\frac{I(t)}{c^2}+o\left(\frac{g_i\vec{x}^2}{2c^2}\right).
\end{align}

We now recall Eq.\eqref{eq:I}, and use the form of the phase shift $\bar{\xi}$ determined numerically. By using the convolution theorem of Fourier integrals, we are able to write:
\begin{align*}
	&I(t)=\left[\int\bar\omega_i \gamma(\vec{\dot{x}}(s)) \sinOb{t\left(1-A(x-B)^2\right)-s} ds\right]_t\\
		&=\biggr[\biggr(\gamma(\vec{\dot{x}}(u))*\sin{\bar\omega_iu}\biggr)\left(t\left(1-A(x-B)^2\right)\right)\biggr]_t=\\
		&=\left[\int\bar\omega_i \gamma\left(\vec{\dot{x}}\left(t\left(1-A(x(t)-B)^2\right)-s\right)\right) \sinOb{s} ds\right]_t
\end{align*}
where $*$ denotes the convolution product.
Now, having to deal with ROs, $\bar\omega_i$ is large, such that we have a fast oscillating trigonometric function.
In the last step, we highlighted that $x(t)$ has to be continuous, i.e. $x\in C^0(\mathbb{R}^{+})$. Furthermore, the function $\vec{\dot{x}}:\mathbb{R}^{+}\to\mathbb{R}^3$ is continuous as well because $\vec{\dot{x}}=\int\{\vec{\ddot{x}}_t\}dt$ where we considered the acceleration of the TP to be stochastic.
Moreover, the TP is a massive particle, meaning that it possesses inertia. It is physically meaningful to assume that the force acting on the TP at each time is upper-bounded, also implying that $\forall t_1,t_2$ $\exists K\in \mathbb{R}^{+}$ such that $\gamma(\vec{\dot{x}}(t_1))-\gamma(\vec{\dot{x}}(t_2))<K$, and it also implies that $\gamma(\vec{\dot{x}}(t))$ is a continuous function.
This allows us to deduce that, for any finite range of times, $\gamma(\vec{\dot{x}}(s))$ is indeed integrable. Then all the hypotheses of the Riemann-Lebesgue (RL) lemma are satisfied, and by application of the RL lemma to the above Fourier integral, we conclude that in the limit $\bar{\omega}_{i}$, we have $I(t)\to 0$, as a direct consequence of the RL lemma.
In reality, of course, $\bar{\omega_{i}}$ is large but finite. Hence, $I(t)$ will not be exactly zero but it will be a small number such that $I(t)/c^{2} \approx 0$ and, therefore, also $\sum_{i}\frac{m_ig_i}{2mc^2}\vec{x}^2(1+I(t))$ is negligible in Eq. \eqref{newterm}.

\subsection{Translation-invariant relativistic Langevin equation}
Having dealt with this last problematic term and then by substituting Eq.\eqref{eq:qRescaled} into the TP EOM, leads us to the relativistic Langevin equation:
\begin{align}
	\begin{split}
		\frac{d}{dt}\Biggr[m\gamma(\dot{\vec{x}})\dot{\vec{x}}\Biggr]=&-\vec{F}_\mathrm{ext}+\vec{F}_p+\\
		&+\int \vec{x}(t-s)\gamma(\vec{\dot{x}}(t-s))K(t,s)ds.
	\end{split}
	\label{eq:lang}
\end{align}
where the conservative force is given by:
\begin{equation}
\vec{F}_\mathrm{ext}=-\nabla (V_\mathrm{ext}-\Phi)
\end{equation}
with $\Phi$ the relativistic renormalization potential derived in the previous section:
\begin{equation}
\Phi=\frac{I(t)}{\gamma(\vec{\dot{x}}(t))} \frac{m_i g_i^{2}}{\bar{\omega}_{i}^{2}} \vec{x}^2(t)
\end{equation}

It is important to notice that the renormalization potential determined above is not just an artificial way of removing an unwanted translation-symmetry breaking term, given by the ``restoring force'' discussed above. The renormalization potential derived above is fully necessary from a physical point of view, as already pointed out by Caldeira and Leggett \cite{CLdiss}. In particular, its physical meaning and significance are as follows. The coupling between the tagged particle and the bath oscillators introduces anharmonicity into both the dynamics of the oscillators and that of the tagged particle. This is because the coupling effectively changes the original conservative potential of the oscillators, as is obvious from e.g. Eq. \eqref{interaction}, and, therefore, it must also change the original potential of the tagged particle. 
In turn, these alterations of the original potential (for both PHB and TP) are inevitably linked to the renormalization of the corresponding natural frequencies, which makes this effect analogous to level-repulsion or avoided-crossing of interacting phonon modes in solids \cite{Klinger,thermoelectric}, as pointed out earlier. In particular, the level-repulsion or avoided-crossing effect becomes larger, meaning that the change in the original eigenfrequencies also becomes larger, for larger values of the coupling constant $g_{i}$. This is a consequence of the fact that larger $g_{i}$ values cause a larger renormalization of the corresponding potentials.

Equation \eqref{eq:lang} is the most important result of this work. By using the technique developed in Ref. \cite{paper}, and by determining the renormalization potential according to the protocol suggested by Caldeira and Leggett \cite{CLdiss}, we managed to arrive at a translation-invariant relativistic Langevin equation.

This equation can also be written in manifestly covariant form, as follows.

Starting from the l.h.s. of Eq. \ref{eq:lang}, by defining the four momentum of the TP as $p^i=(p^0,\vec{p})=(E/c,\vec{\dot{x}}m\gamma(\vec{\dot{x}}))$ we get the covariant form \cite{relBrown1+1}
\begin{equation}
		\frac{d}{dt}\left[\gamma(\vec{\dot{x}})m \vec{\dot{x}}\right]=\frac{d}{dt}p^i
\end{equation}

For the r.h.s. things get a little trickier because of the stochastic force, but first let us proceed in order: both $\vec{F}'_r$ and $\vec{F}_{\mathrm{ext}_b}$ can be made into spatial components of four-vector whose temporal components are equal to zero by construction.

Let us now deal with the stochastic force, and define a covariant rank-2 tensor, i.e. the memory tensor $\tensor{K}{^\mu}_{\nu}(t,s)=\mathrm{diag}(0,K'^1(t,s),K'^2(t,s),K'^3(t,s))$. We note that in Eq. \ref{eq:lang}, extending the integrand to the four-vector formalism, we can rewrite $\gamma(\vec{\dot{x}}(t-s))\vec{\dot{x}}(t-s)=p\indices{^\mu}(t-s)/m$.
Using these two devices and the Einstein summation notation, we finally arrive at:
\begin{equation}
		\frac{d}{dt}p\indices{^\mu}=-F_\mathrm{ext}^\mu+F_p^\mu-\frac{1}{m}\int_0^t\tensor{K}{^\mu}_{\nu}(t,s) p\indices{^\nu}(t-s)ds
		\label{eq:covlang}
\end{equation}
which is the fully covariant form of the relativistic Langevin equation we just derived from first principles. 
We used Greek indices as is customary for four-vectors, whereas earlier in the paper we used Latin indices for the spatial components only. 
Moreover, another interesting thing to point out is that we are still using $t$ as our coordinate time, this can be done because we are working in the instantaneous rest frame of the TP hence we could exchange $t$ with the propert time $\tau$. Alternatively, we could have defined another parameter, as it is sometimes done in the literature on general and special relativity (e.g. in \cite{goldstein} the parameter $\lambda$ is used).
Furthermore, by following this procedure, we can see that we are working under the assumption that stochastic behaviour does not affect the time component, as prescribed by $\dot{p}^0=0$. 
This is indeed quite standard, and agrees with previous models in the literature \cite{debbasch1997}.

\section{Relativistic fluctuation-dissipation theorem}
We shall consider the implications of the above results on the fluctuation-dissipation theorem (FDT) associated with the relativistic Langevin equation.
Let us start by noting that we can treat the dependence on $\bar\omega$ of $\bar\xi$ as an explicit dependence $\bar\xi(t,\bar\omega_i)$, which allows us to define a density of states $\rho(\bar\omega)$, where  $\bar\omega_i$ is promoted to a continuous variable.
With this passage to a continuous frequency spectrum, we can use Fourier transforms to write all the various terms that form the relativistic Langeving equation, as follows:
\begin{align}
K(t,s)&=\int_0^\infty d\bar\omega\rho(\bar\omega)\frac{g(\bar\omega)^2}{\bar\omega^2}\gamma^{-1}(\dot{\vec{x}}(s))\times\nonumber\\
&\times\left[\int dy \gamma(\dot{\vec{x}}(y))\sin\left(\bar\omega\left( t-y-\frac{\bar\xi_i(y,\bar\omega)}{c}\right)\right)\right]_{y=s} \label{eq:relkernel_cont}
\end{align}
where the information carried by the subscript $i$ resides now in the explicit dependence on the continuous variable $\bar\omega$.

The above form for the memory kernel is of great interest as we can draw a parallel between the classical framework and the relativistic one. 
In the non-relativistic case, one has \cite{zwanzig}:
\begin{equation}
K(t)=\int_0^\infty d\omega\rho(\omega)\frac{g(\omega)^2}{\omega^2}\cos(\omega t)\label{eq:kernel_class}
\end{equation}
In particular, Zwanzig showed that in the case where $\rho(\bar\omega)=\alpha\bar\omega^2$, as for bosonic oscillators, and $g(\bar\omega)=\mathrm{const}$, then in the non-relativistic limit we would have $K(t)\propto\delta(t)$ by evaluating the above integral in Eq. \eqref{eq:kernel_class}  \cite{zwanzig}.
By assuming that the system is at thermodynamic equilibrium and thus follows the Maxwell-Boltzmann (or Maxwell-J \"uttner) distribution at $t=0$ (to evaluate the ensemble average of the stochastic force auto-correlation $\langle \vec{F}'_p(t) \vec{F}'_p(t')$), leads then to the well-known Markovian FDT: $\langle \vec{F}'_p(t) \vec{F}'_p(t') \rangle = m k_B T \delta(t-t')$, as derived by Zwanzig \cite{zwanzig}.

In the relativistic case, this simplification is no longer possible, as we have a dependency on the oscillator's trajectory via $\bar\xi$. To solve this problem we can substitute into Eq. \eqref{eq:relkernel_cont} the form Eq. \eqref{eq:xi} deduced from the numerics. As shown in Ref. \cite{paper}, this makes it possible for the memory kernel to be proportional to $\delta(t)$ even in this framework.
However, the assumption $g(\bar\omega)=\mathrm{const}$ is untenable in the relativistic case. This is because this assumption is tantamount to saying that the TP interacts with the same strength with all the ROs in the system, irrespective of their separation in space-time from the TP. This is obviously in contradiction with the principle of locality, which lies at the heart of special relativity.

Following the same strategy as in \cite{paper}, and assuming that $g(\bar\omega)\neq \mathrm{const}$ in compliance with the principle of locality in special relativity, leads us to the relativistic FDT in the following form:
\begin{align}
		\langle \vec{F}'_p(t) \vec{F}'_p(t') \rangle = m k_B T K(t-t')
\end{align}
where $K(t-t')\neq \delta(t-t')$ and $K(t-t')$ is specified by Eq. \eqref{eq:relkernel_cont}. This implies that the relativistic FDT is, in general, non-Markovian and the assumption of a Markovian FDT in relativistic systems is unjustified. This provides a first-principles justification to the observation that memory and non-Markovianity are typically observed in the QGP (barring the regime of lowest energies), see e.g. \cite{Brambilla,Datta_2022}, where lattice QCD calculations indicate that the spectral function is non-Ohmic, i.e. non-Markovian, at high enough energies.

\section{Conclusion}

In summary, we considered the derivation of the relativistic Langevin equation from a particle-bath Lagrangian proposed in Ref. \cite{paper} and we applied the protocol devised by Caldeira and Leggett \cite{CLdiss} to determine the renormalization potential for the tagged particle.
This renormalization potential is important because, on one hand, it is physically necessary as the coupling to the oscillators bath necessarily renormalizes the conservative potential acting on the tagged particle. Furthermore, the introduction of the renormalization potential is necessary to derive a fully translation-invariant relativistic Langevin equation from first principles, which has been done here for the first time.
The form of the relativistic renormalization potential obtained here, given by  $\Phi=\frac{I(t)}{\gamma(\vec{\dot{x}}(t))} \frac{m_i g_i^{2}}{\bar{\omega}_{i}^{2}} \vec{x}^2(t)$ reduces to the classical one in the non-relativistic limit, i.e. to $\Phi= \frac{m_i g_i^{2}}{\bar{\omega}_{i}^{2}} \vec{x}^2(t)$ (that was already obtained by Caldeira and Leggett themselves \cite{paper}).

These results also further establish the non-Markovianity of the relativistic fluctuation-dissipation theorem which is associated with the relativistic Langevin equation. In other words, special relativity necessarily introduces non-Markovian corrections to the noise.
In future work, the form of the relativistic memory kernel $K(t-t')$ as given by Eq. \eqref{eq:relkernel_cont}, will be studied for simple cases, although this requires numerical work to implement trajectories of both the tagged particle and the relativistic oscillators. From the generated trajectories, also the renormalization potential $\Phi$ can then be computed. It is hoped that the numerical calculations will lead to analytical parametrizations for the memory kernel that can be implemented in hydrodynamic simulations. The form of the kernel can also be compared with non-Markovian spectral densities measured in lattice QCD simulations of the QGP \cite{Brambilla,Datta_2022}.

In future work, these results can be used as the starting point for the mathematical modelling of relativistic fluids, weakly relativistic electrons in graphene and other topological materials, relativistic plasmas and for further generalizations to high-energy density matter. For example, we expect the above relativistic fluctuation-dissipation theorem to be highly relevant for the Johnson-Nyquist noise of weakly relativistic electrons in graphene, where the form of the fluctuation-dissipation theorem determines the statistics of the noise \cite{Skinner}.

\subsection*{Acknowledgments} 
Alessio Zaccone gratefully acknowledges funding from the European Union through Horizon Europe ERC Grant number: 101043968 ``Multimech'', and from US Army Research Office through contract nr.W911NF-22-2-0256.

\appendix
\section{Physical justification of $I(t)/c^2 \ll 1$}
\label{app:approx}
Here we will provide a more physical justification for the key argument made in the context of Eq.\eqref{eq:approx}.

Let us start by using the assumption of a massive TP, which, as such it will never reach the speed of light, meaning that we can take $\norm{\vec{\dot{x}}(t)}_3=c\mathcal{K}(t)$ with $\mathcal{K}:\mathbb{R}^+\to[0,1)$.

Now we can consider two scenarios, the first one being of a dissipative adiabatic system, meaning that the TP loses energy to the PHB.
The second scenario assumes a finite observation time and a finite number of PHB oscillators, allowing us to have a constraint on the maximum speed of the TP.
\\

Assuming a dissipative behavior of the PHB enables us to write that the TP energy $E_{\mathrm{TP}}(t)^{2}<E_{\mathrm{TP}}(0)^2$ $\forall t$ as work is exchanged between the TP and the heat bath modes.
However, based on the initial conditions of the problem $E_{\mathrm{TP}}(0)$ is a known quantity, so by using the definition of relativistic energy and introducing $\mathcal{K}(t)$ we can write:
\begin{align*}
	E_{\mathrm{TP}}(t)^2&<E_{\mathrm{TP}}(0)^2\\
	(\norm{p(t)}_4c)^2&<E_{\mathrm{TP}}(0)^2-E_0^2\\
	\gamma(\vec{\dot{x}}(t))\norm{\vec{\dot{x}}(t)}_3&<\sqrt{\frac{E_{\mathrm{TP}}(0)^2-E_0^2}{m^2c^2}}\\
	\frac{\mathcal{K}}{\sqrt{1-\mathcal{K}^2}}&<\frac{1}{c}\sqrt{\frac{E_{\mathrm{TP}}(0)^2}{E_0^2}-1}=A
\end{align*}
then $\mathcal{K}(t)\in[0,\sqrt{A/(A+1)}]$ and indeed the upper bound is less than $1$.
Now that $\mathcal{K}(t)$ has an upper bound, we use it to control the integral:
\begin{align*}
	I(t)&<\gamma\left(c\sqrt{\frac{A}{A+1}}\right)\left[\int\sinOb{t-s-\frac{\bar{\xi}_i(t)}{c}} ds\right]_t<\\
	&<\sqrt{(A+1)}\left[\int 1 ds\right]_t=\sqrt{A+1}t
\end{align*}
from which it follows that $I(t)/c^2\ll 1$ is a sensible approximation indeed.
\\

The second approach gives more context about the maximum expected velocities that the TP can achieve. This is done by assuming that our system is allowed to evolve only for a finite time window or observation time.
Having a finite number of PHB constituents means that the force acting on the TP is indeed finite, coupling it with the finite time of observation allows us to say that there exists an upper bound to the magnitude of the TP speed during the observation period, such that $\exists\bar{ \mathcal{K}}=\mathrm{max(\mathcal{K}(t))}$.

Now we can control the integral as we did before, this time obtaining $I(t)<\gamma(\bar{\mathcal{K}} c)t$.
Hence we get $I(t)\ll c^2$, or, in a more complete form, $t^2\ll c^2(1-\bar{\mathcal{K}}^2)$.
To compare the order of magnitudes, let us assume that $t\approx10^n$ then $\bar{\mathcal{K}}< \sqrt{1-10^{-14-2n}}$ which for a reasonable amount of time $t<10^{1}$ equates to a Lorentz factor of up to $\gamma\approx10^6$, meaning that this approximation is indeed sensible for all phenomena occuring in between the relativistic and the ultrarelativistic limit.

\newpage
\bibliographystyle{apsrev4-1}
\bibliography{ref}

\begin{thebibliography}{36}%
\makeatletter
\providecommand \@ifxundefined [1]{%
 \@ifx{#1\undefined}
}%
\providecommand \@ifnum [1]{%
 \ifnum #1\expandafter \@firstoftwo
 \else \expandafter \@secondoftwo
 \fi
}%
\providecommand \@ifx [1]{%
 \ifx #1\expandafter \@firstoftwo
 \else \expandafter \@secondoftwo
 \fi
}%
\providecommand \natexlab [1]{#1}%
\providecommand \enquote  [1]{``#1''}%
\providecommand \bibnamefont  [1]{#1}%
\providecommand \bibfnamefont [1]{#1}%
\providecommand \citenamefont [1]{#1}%
\providecommand \href@noop [0]{\@secondoftwo}%
\providecommand \href [0]{\begingroup \@sanitize@url \@href}%
\providecommand \@href[1]{\@@startlink{#1}\@@href}%
\providecommand \@@href[1]{\endgroup#1\@@endlink}%
\providecommand \@sanitize@url [0]{\catcode `\\12\catcode `\$12\catcode
  `\&12\catcode `\#12\catcode `\^12\catcode `\_12\catcode `\%12\relax}%
\providecommand \@@startlink[1]{}%
\providecommand \@@endlink[0]{}%
\providecommand \url  [0]{\begingroup\@sanitize@url \@url }%
\providecommand \@url [1]{\endgroup\@href {#1}{\urlprefix }}%
\providecommand \urlprefix  [0]{URL }%
\providecommand \Eprint [0]{\href }%
\providecommand \doibase [0]{http://dx.doi.org/}%
\providecommand \selectlanguage [0]{\@gobble}%
\providecommand \bibinfo  [0]{\@secondoftwo}%
\providecommand \bibfield  [0]{\@secondoftwo}%
\providecommand \translation [1]{[#1]}%
\providecommand \BibitemOpen [0]{}%
\providecommand \bibitemStop [0]{}%
\providecommand \bibitemNoStop [0]{.\EOS\space}%
\providecommand \EOS [0]{\spacefactor3000\relax}%
\providecommand \BibitemShut  [1]{\csname bibitem#1\endcsname}%
\let\auto@bib@innerbib\@empty
\bibitem [{\citenamefont {Ornstein}(1930)}]{langToDiff}%
  \BibitemOpen
  \bibfield  {author} {\bibinfo {author} {\bibfnamefont {L.~S.}\ \bibnamefont
  {Ornstein}},\ }\href@noop {} {\bibfield  {journal} {\bibinfo  {journal}
  {Physical review}\ }\textbf {\bibinfo {volume} {36}},\ \bibinfo {pages} {823}
  (\bibinfo {year} {1930})}\BibitemShut {NoStop}%
\bibitem [{\citenamefont {Zwanzig}(2001)}]{zwanzig}%
  \BibitemOpen
  \bibfield  {author} {\bibinfo {author} {\bibfnamefont {R.}~\bibnamefont
  {Zwanzig}},\ }\href@noop {} {\emph {\bibinfo {title} {Nonequilibrium
  statistical mechanics}}}\ (\bibinfo  {publisher} {Oxford University press},\
  \bibinfo {address} {New York},\ \bibinfo {year} {2001})\BibitemShut {NoStop}%
\bibitem [{\citenamefont {Beck}\ and\ \citenamefont
  {Roepstorff}(1990)}]{langToHydro}%
  \BibitemOpen
  \bibfield  {author} {\bibinfo {author} {\bibfnamefont {C.}~\bibnamefont
  {Beck}}\ and\ \bibinfo {author} {\bibfnamefont {G.}~\bibnamefont
  {Roepstorff}},\ }\href {\doibase
  https://doi.org/10.1016/0378-4371(90)90195-X} {\bibfield  {journal} {\bibinfo
   {journal} {Physica A: Statistical Mechanics and its Applications}\ }\textbf
  {\bibinfo {volume} {165}},\ \bibinfo {pages} {270} (\bibinfo {year}
  {1990})}\BibitemShut {NoStop}%
\bibitem [{\citenamefont {Bu}\ \emph {et~al.}(2022)\citenamefont {Bu},
  \citenamefont {Zhang},\ and\ \citenamefont {Zhang}}]{Zhang}%
  \BibitemOpen
  \bibfield  {author} {\bibinfo {author} {\bibfnamefont {Y.}~\bibnamefont
  {Bu}}, \bibinfo {author} {\bibfnamefont {B.}~\bibnamefont {Zhang}}, \ and\
  \bibinfo {author} {\bibfnamefont {J.}~\bibnamefont {Zhang}},\ }\href
  {\doibase 10.1103/PhysRevD.106.086014} {\bibfield  {journal} {\bibinfo
  {journal} {Phys. Rev. D}\ }\textbf {\bibinfo {volume} {106}},\ \bibinfo
  {pages} {086014} (\bibinfo {year} {2022})}\BibitemShut {NoStop}%
\bibitem [{\citenamefont {Mullins}\ \emph {et~al.}(2023)\citenamefont
  {Mullins}, \citenamefont {Hippert}, \citenamefont {Gavassino},\ and\
  \citenamefont {Noronha}}]{mullins2023relativistic}%
  \BibitemOpen
  \bibfield  {author} {\bibinfo {author} {\bibfnamefont {N.}~\bibnamefont
  {Mullins}}, \bibinfo {author} {\bibfnamefont {M.}~\bibnamefont {Hippert}},
  \bibinfo {author} {\bibfnamefont {L.}~\bibnamefont {Gavassino}}, \ and\
  \bibinfo {author} {\bibfnamefont {J.}~\bibnamefont {Noronha}},\ }\href@noop
  {} {\bibfield  {journal} {\bibinfo  {journal} {arXiv preprint
  arXiv:2309.00512}\ } (\bibinfo {year} {2023})}\BibitemShut {NoStop}%
\bibitem [{\citenamefont {Singh}\ \emph {et~al.}(2019)\citenamefont {Singh},
  \citenamefont {Shen}, \citenamefont {McDonald}, \citenamefont {Jeon},\ and\
  \citenamefont {Gale}}]{Singh}%
  \BibitemOpen
  \bibfield  {author} {\bibinfo {author} {\bibfnamefont {M.}~\bibnamefont
  {Singh}}, \bibinfo {author} {\bibfnamefont {C.}~\bibnamefont {Shen}},
  \bibinfo {author} {\bibfnamefont {S.}~\bibnamefont {McDonald}}, \bibinfo
  {author} {\bibfnamefont {S.}~\bibnamefont {Jeon}}, \ and\ \bibinfo {author}
  {\bibfnamefont {C.}~\bibnamefont {Gale}},\ }\href {\doibase
  https://doi.org/10.1016/j.nuclphysa.2018.10.061} {\bibfield  {journal}
  {\bibinfo  {journal} {Nuclear Physics A}\ }\textbf {\bibinfo {volume}
  {982}},\ \bibinfo {pages} {319} (\bibinfo {year} {2019})},\ \bibinfo {note}
  {the 27th International Conference on Ultrarelativistic Nucleus-Nucleus
  Collisions: Quark Matter 2018}\BibitemShut {NoStop}%
\bibitem [{\citenamefont {Batini}\ \emph {et~al.}(2023)\citenamefont {Batini},
  \citenamefont {Grossi},\ and\ \citenamefont {Wink}}]{batini}%
  \BibitemOpen
  \bibfield  {author} {\bibinfo {author} {\bibfnamefont {L.}~\bibnamefont
  {Batini}}, \bibinfo {author} {\bibfnamefont {E.}~\bibnamefont {Grossi}}, \
  and\ \bibinfo {author} {\bibfnamefont {N.}~\bibnamefont {Wink}},\ }\href@noop
  {} {\enquote {\bibinfo {title} {Dissipation dynamics of a scalar field},}\ }
  (\bibinfo {year} {2023}),\ \Eprint {http://arxiv.org/abs/2309.06586}
  {arXiv:2309.06586 [hep-th]} \BibitemShut {NoStop}%
\bibitem [{\citenamefont {Moore}\ and\ \citenamefont
  {Teaney}(2005)}]{Teaney_2005}%
  \BibitemOpen
  \bibfield  {author} {\bibinfo {author} {\bibfnamefont {G.~D.}\ \bibnamefont
  {Moore}}\ and\ \bibinfo {author} {\bibfnamefont {D.}~\bibnamefont {Teaney}},\
  }\href {\doibase 10.1103/PhysRevC.71.064904} {\bibfield  {journal} {\bibinfo
  {journal} {Phys. Rev. C}\ }\textbf {\bibinfo {volume} {71}},\ \bibinfo
  {pages} {064904} (\bibinfo {year} {2005})}\BibitemShut {NoStop}%
\bibitem [{\citenamefont {Petreczky}\ and\ \citenamefont
  {Teaney}(2006)}]{Teaney_2006}%
  \BibitemOpen
  \bibfield  {author} {\bibinfo {author} {\bibfnamefont {P.}~\bibnamefont
  {Petreczky}}\ and\ \bibinfo {author} {\bibfnamefont {D.}~\bibnamefont
  {Teaney}},\ }\href {\doibase 10.1103/PhysRevD.73.014508} {\bibfield
  {journal} {\bibinfo  {journal} {Phys. Rev. D}\ }\textbf {\bibinfo {volume}
  {73}},\ \bibinfo {pages} {014508} (\bibinfo {year} {2006})}\BibitemShut
  {NoStop}%
\bibitem [{\citenamefont {He}\ \emph {et~al.}(2023)\citenamefont {He},
  \citenamefont {van Hees},\ and\ \citenamefont {Rapp}}]{He_2023}%
  \BibitemOpen
  \bibfield  {author} {\bibinfo {author} {\bibfnamefont {M.}~\bibnamefont
  {He}}, \bibinfo {author} {\bibfnamefont {H.}~\bibnamefont {van Hees}}, \ and\
  \bibinfo {author} {\bibfnamefont {R.}~\bibnamefont {Rapp}},\ }\href {\doibase
  10.1016/j.ppnp.2023.104020} {\bibfield  {journal} {\bibinfo  {journal}
  {Progress in Particle and Nuclear Physics}\ }\textbf {\bibinfo {volume}
  {130}},\ \bibinfo {pages} {104020} (\bibinfo {year} {2023})}\BibitemShut
  {NoStop}%
\bibitem [{\citenamefont {Win}\ \emph {et~al.}(2023)\citenamefont {Win},
  \citenamefont {Aung}, \citenamefont {Khandal},\ and\ \citenamefont
  {Ghosh}}]{win2023graphene}%
  \BibitemOpen
  \bibfield  {author} {\bibinfo {author} {\bibfnamefont {T.~Z.}\ \bibnamefont
  {Win}}, \bibinfo {author} {\bibfnamefont {C.~W.}\ \bibnamefont {Aung}},
  \bibinfo {author} {\bibfnamefont {G.}~\bibnamefont {Khandal}}, \ and\
  \bibinfo {author} {\bibfnamefont {S.}~\bibnamefont {Ghosh}},\ }\href@noop {}
  {\enquote {\bibinfo {title} {Graphene is neither relativistic nor
  non-relativistic case: Thermodynamics aspects},}\ } (\bibinfo {year}
  {2023}),\ \Eprint {http://arxiv.org/abs/2307.05395} {arXiv:2307.05395
  [cond-mat.str-el]} \BibitemShut {NoStop}%
\bibitem [{\citenamefont {Krishna~Kumar}(2017)}]{Levitov}%
  \BibitemOpen
  \bibfield  {author} {\bibinfo {author} {\bibfnamefont {R.~e.~a.}\
  \bibnamefont {Krishna~Kumar}},\ }\href {https://doi.org/10.1038/nphys4240}
  {\bibfield  {journal} {\bibinfo  {journal} {Nature Physics}\ }\textbf
  {\bibinfo {volume} {13}},\ \bibinfo {pages} {1182} (\bibinfo {year}
  {2017})}\BibitemShut {NoStop}%
\bibitem [{\citenamefont {Debbasch}\ \emph {et~al.}(1997)\citenamefont
  {Debbasch}, \citenamefont {Mallick},\ and\ \citenamefont
  {Rivet}}]{debbasch1997}%
  \BibitemOpen
  \bibfield  {author} {\bibinfo {author} {\bibfnamefont {F.}~\bibnamefont
  {Debbasch}}, \bibinfo {author} {\bibfnamefont {K.}~\bibnamefont {Mallick}}, \
  and\ \bibinfo {author} {\bibfnamefont {J.~P.}\ \bibnamefont {Rivet}},\
  }\href@noop {} {\bibfield  {journal} {\bibinfo  {journal} {Journal of
  statistical physics}\ }\textbf {\bibinfo {volume} {88}},\ \bibinfo {pages}
  {945} (\bibinfo {year} {1997})}\BibitemShut {NoStop}%
\bibitem [{\citenamefont {Koide}\ \emph {et~al.}(2007)\citenamefont {Koide},
  \citenamefont {Denicol}, \citenamefont {Mota},\ and\ \citenamefont
  {Kodama}}]{koide}%
  \BibitemOpen
  \bibfield  {author} {\bibinfo {author} {\bibfnamefont {T.}~\bibnamefont
  {Koide}}, \bibinfo {author} {\bibfnamefont {G.~S.}\ \bibnamefont {Denicol}},
  \bibinfo {author} {\bibfnamefont {P.}~\bibnamefont {Mota}}, \ and\ \bibinfo
  {author} {\bibfnamefont {T.}~\bibnamefont {Kodama}},\ }\href {\doibase
  10.1103/PhysRevC.75.034909} {\bibfield  {journal} {\bibinfo  {journal} {Phys.
  Rev. C}\ }\textbf {\bibinfo {volume} {75}},\ \bibinfo {pages} {034909}
  (\bibinfo {year} {2007})}\BibitemShut {NoStop}%
\bibitem [{\citenamefont {Dunkel}\ and\ \citenamefont
  {Hänggi}(2005)}]{relBrown1+1}%
  \BibitemOpen
  \bibfield  {author} {\bibinfo {author} {\bibfnamefont {J.}~\bibnamefont
  {Dunkel}}\ and\ \bibinfo {author} {\bibfnamefont {P.}~\bibnamefont
  {Hänggi}},\ }\href@noop {} {\bibfield  {journal} {\bibinfo  {journal}
  {Physical review. E, Statistical, nonlinear, and soft matter physics}\
  }\textbf {\bibinfo {volume} {71}},\ \bibinfo {pages} {016124} (\bibinfo
  {year} {2005})}\BibitemShut {NoStop}%
\bibitem [{\citenamefont {Petrosyan}\ and\ \citenamefont
  {Zaccone}(2022)}]{paper}%
  \BibitemOpen
  \bibfield  {author} {\bibinfo {author} {\bibfnamefont {A.}~\bibnamefont
  {Petrosyan}}\ and\ \bibinfo {author} {\bibfnamefont {A.}~\bibnamefont
  {Zaccone}},\ }\href {\doibase 10.1088/1751-8121/ac3a33} {\bibfield  {journal}
  {\bibinfo  {journal} {Journal of Physics A: Mathematical and Theoretical}\
  }\textbf {\bibinfo {volume} {55}},\ \bibinfo {pages} {015001} (\bibinfo
  {year} {2022})}\BibitemShut {NoStop}%
\bibitem [{\citenamefont {Debbasch}\ \emph {et~al.}(2012)\citenamefont
  {Debbasch}, \citenamefont {Espaze}, \citenamefont {Foulonneau},\ and\
  \citenamefont {Rivet}}]{debbasch1998}%
  \BibitemOpen
  \bibfield  {author} {\bibinfo {author} {\bibfnamefont {F.}~\bibnamefont
  {Debbasch}}, \bibinfo {author} {\bibfnamefont {D.}~\bibnamefont {Espaze}},
  \bibinfo {author} {\bibfnamefont {V.}~\bibnamefont {Foulonneau}}, \ and\
  \bibinfo {author} {\bibfnamefont {J.-P.}\ \bibnamefont {Rivet}},\ }\href@noop
  {} {\bibfield  {journal} {\bibinfo  {journal} {Physica A}\ }\textbf {\bibinfo
  {volume} {391}},\ \bibinfo {pages} {3797} (\bibinfo {year}
  {2012})}\BibitemShut {NoStop}%
\bibitem [{\citenamefont {BARBACHOUX}\ \emph
  {et~al.}(2001{\natexlab{a}})\citenamefont {BARBACHOUX}, \citenamefont
  {DEBBASCH},\ and\ \citenamefont {RIVET}}]{debbasch2001}%
  \BibitemOpen
  \bibfield  {author} {\bibinfo {author} {\bibfnamefont {C.}~\bibnamefont
  {BARBACHOUX}}, \bibinfo {author} {\bibfnamefont {F.}~\bibnamefont
  {DEBBASCH}}, \ and\ \bibinfo {author} {\bibfnamefont {J.~P.}\ \bibnamefont
  {RIVET}},\ }\href@noop {} {\bibfield  {journal} {\bibinfo  {journal} {The
  European physical journal. B, Condensed matter physics}\ }\textbf {\bibinfo
  {volume} {19}},\ \bibinfo {pages} {37} (\bibinfo {year}
  {2001}{\natexlab{a}})}\BibitemShut {NoStop}%
\bibitem [{\citenamefont {BARBACHOUX}\ \emph
  {et~al.}(2001{\natexlab{b}})\citenamefont {BARBACHOUX}, \citenamefont
  {DEBBASCH},\ and\ \citenamefont {RIVET}}]{debbasch2001_2}%
  \BibitemOpen
  \bibfield  {author} {\bibinfo {author} {\bibfnamefont {C.}~\bibnamefont
  {BARBACHOUX}}, \bibinfo {author} {\bibfnamefont {F.}~\bibnamefont
  {DEBBASCH}}, \ and\ \bibinfo {author} {\bibfnamefont {J.~P.}\ \bibnamefont
  {RIVET}},\ }\href@noop {} {\bibfield  {journal} {\bibinfo  {journal} {The
  European physical journal. B, Condensed matter physics}\ }\textbf {\bibinfo
  {volume} {23}},\ \bibinfo {pages} {487} (\bibinfo {year}
  {2001}{\natexlab{b}})}\BibitemShut {NoStop}%
\bibitem [{\citenamefont {Dunkel}\ and\ \citenamefont
  {Hänggi}(2009)}]{dunkel2009}%
  \BibitemOpen
  \bibfield  {author} {\bibinfo {author} {\bibfnamefont {J.}~\bibnamefont
  {Dunkel}}\ and\ \bibinfo {author} {\bibfnamefont {P.}~\bibnamefont
  {Hänggi}},\ }\href {\doibase https://doi.org/10.1016/j.physrep.2008.12.001}
  {\bibfield  {journal} {\bibinfo  {journal} {Physics Reports}\ }\textbf
  {\bibinfo {volume} {471}},\ \bibinfo {pages} {1} (\bibinfo {year}
  {2009})}\BibitemShut {NoStop}%
\bibitem [{\citenamefont {Cattaneo}(1948)}]{cattaneo}%
  \BibitemOpen
  \bibfield  {author} {\bibinfo {author} {\bibfnamefont {C.}~\bibnamefont
  {Cattaneo}},\ }\href@noop {} {\bibfield  {journal} {\bibinfo  {journal} {Atti
  Semin. Mat. Fis. della Università di Modena}\ }\textbf {\bibinfo {volume}
  {3}} (\bibinfo {year} {1948})}\BibitemShut {NoStop}%
\bibitem [{\citenamefont {Brezis}(2010)}]{brezis}%
  \BibitemOpen
  \bibfield  {author} {\bibinfo {author} {\bibfnamefont {H.}~\bibnamefont
  {Brezis}},\ }\href {https://books.google.it/books?id=GAA2XqOIIGoC} {\emph
  {\bibinfo {title} {Functional Analysis, Sobolev Spaces and Partial
  Differential Equations}}},\ Universitext\ (\bibinfo  {publisher} {Springer
  New York},\ \bibinfo {year} {2010})\BibitemShut {NoStop}%
\bibitem [{\citenamefont {Cahill}(2019)}]{cahill}%
  \BibitemOpen
  \bibfield  {author} {\bibinfo {author} {\bibfnamefont {K.}~\bibnamefont
  {Cahill}},\ }\href@noop {} {\emph {\bibinfo {title} {Physical
  mathematics}}},\ \bibinfo {edition} {2nd}\ ed.\ (\bibinfo  {publisher}
  {Cambridge University press},\ \bibinfo {address} {Cambridge [etc},\ \bibinfo
  {year} {2019})\BibitemShut {NoStop}%
\bibitem [{\citenamefont {L.~D.~Landau}(1969)}]{lifshitz}%
  \BibitemOpen
  \bibfield  {author} {\bibinfo {author} {\bibfnamefont {I.~L.}\ \bibnamefont
  {L.~D.~Landau}},\ }\href@noop {} {\emph {\bibinfo {title} {Mechanics}}}\
  (\bibinfo  {publisher} {Pergamon press},\ \bibinfo {address} {Oxford},\
  \bibinfo {year} {1969})\BibitemShut {NoStop}%
\bibitem [{\citenamefont {MacColl}(1957)}]{teo_RHO}%
  \BibitemOpen
  \bibfield  {author} {\bibinfo {author} {\bibfnamefont {L.~A.}\ \bibnamefont
  {MacColl}},\ }\href {\doibase 10.1119/1.1934543} {\bibfield  {journal}
  {\bibinfo  {journal} {American Journal of Physics}\ }\textbf {\bibinfo
  {volume} {25}},\ \bibinfo {pages} {535} (\bibinfo {year} {1957})},\ \Eprint
  {http://arxiv.org/abs/https://doi.org/10.1119/1.1934543}
  {https://doi.org/10.1119/1.1934543} \BibitemShut {NoStop}%
\bibitem [{\citenamefont {Xu}\ \emph {et~al.}(2008)\citenamefont {Xu},
  \citenamefont {Wang}, \citenamefont {Duan}, \citenamefont {Gu},\ and\
  \citenamefont {Li}}]{Bao}%
  \BibitemOpen
  \bibfield  {author} {\bibinfo {author} {\bibfnamefont {Y.}~\bibnamefont
  {Xu}}, \bibinfo {author} {\bibfnamefont {J.-S.}\ \bibnamefont {Wang}},
  \bibinfo {author} {\bibfnamefont {W.}~\bibnamefont {Duan}}, \bibinfo {author}
  {\bibfnamefont {B.-L.}\ \bibnamefont {Gu}}, \ and\ \bibinfo {author}
  {\bibfnamefont {B.}~\bibnamefont {Li}},\ }\href {\doibase
  10.1103/PhysRevB.78.224303} {\bibfield  {journal} {\bibinfo  {journal} {Phys.
  Rev. B}\ }\textbf {\bibinfo {volume} {78}},\ \bibinfo {pages} {224303}
  (\bibinfo {year} {2008})}\BibitemShut {NoStop}%
\bibitem [{\citenamefont {Casella}\ and\ \citenamefont
  {Zaccone}(2021)}]{Casella}%
  \BibitemOpen
  \bibfield  {author} {\bibinfo {author} {\bibfnamefont {L.}~\bibnamefont
  {Casella}}\ and\ \bibinfo {author} {\bibfnamefont {A.}~\bibnamefont
  {Zaccone}},\ }\href {\doibase 10.1088/1361-648X/abdb68} {\bibfield  {journal}
  {\bibinfo  {journal} {Journal of Physics: Condensed Matter}\ }\textbf
  {\bibinfo {volume} {33}},\ \bibinfo {pages} {165401} (\bibinfo {year}
  {2021})}\BibitemShut {NoStop}%
\bibitem [{\citenamefont {Caldeira}\ and\ \citenamefont
  {Leggett}(1983)}]{CLdiss}%
  \BibitemOpen
  \bibfield  {author} {\bibinfo {author} {\bibfnamefont {A.}~\bibnamefont
  {Caldeira}}\ and\ \bibinfo {author} {\bibfnamefont {A.}~\bibnamefont
  {Leggett}},\ }\href {\doibase https://doi.org/10.1016/0003-4916(83)90202-6}
  {\bibfield  {journal} {\bibinfo  {journal} {Annals of Physics}\ }\textbf
  {\bibinfo {volume} {149}},\ \bibinfo {pages} {374} (\bibinfo {year}
  {1983})}\BibitemShut {NoStop}%
\bibitem [{\citenamefont {Gottwald}\ \emph {et~al.}(2016)\citenamefont
  {Gottwald}, \citenamefont {Ivanov},\ and\ \citenamefont
  {Kühn}}]{anharmonic}%
  \BibitemOpen
  \bibfield  {author} {\bibinfo {author} {\bibfnamefont {F.}~\bibnamefont
  {Gottwald}}, \bibinfo {author} {\bibfnamefont {S.~D.}\ \bibnamefont
  {Ivanov}}, \ and\ \bibinfo {author} {\bibfnamefont {O.}~\bibnamefont
  {Kühn}},\ }\href {\doibase 10.1063/1.4946872} {\bibfield  {journal}
  {\bibinfo  {journal} {The Journal of Chemical Physics}\ }\textbf {\bibinfo
  {volume} {144}},\ \bibinfo {pages} {164102} (\bibinfo {year} {2016})},\
  \Eprint {http://arxiv.org/abs/https://doi.org/10.1063/1.4946872}
  {https://doi.org/10.1063/1.4946872} \BibitemShut {NoStop}%
\bibitem [{\citenamefont {Petruccione}\ and\ \citenamefont
  {Vacchini}(2005)}]{petruccione2005}%
  \BibitemOpen
  \bibfield  {author} {\bibinfo {author} {\bibfnamefont {F.}~\bibnamefont
  {Petruccione}}\ and\ \bibinfo {author} {\bibfnamefont {B.}~\bibnamefont
  {Vacchini}},\ }\href {\doibase 10.1103/PhysRevE.71.046134} {\bibfield
  {journal} {\bibinfo  {journal} {Phys. Rev. E}\ }\textbf {\bibinfo {volume}
  {71}},\ \bibinfo {pages} {046134} (\bibinfo {year} {2005})}\BibitemShut
  {NoStop}%
\bibitem [{\citenamefont {Baggioli}\ \emph {et~al.}(2019)\citenamefont
  {Baggioli}, \citenamefont {Cui},\ and\ \citenamefont
  {Zaccone}}]{thermoelectric}%
  \BibitemOpen
  \bibfield  {author} {\bibinfo {author} {\bibfnamefont {M.}~\bibnamefont
  {Baggioli}}, \bibinfo {author} {\bibfnamefont {B.}~\bibnamefont {Cui}}, \
  and\ \bibinfo {author} {\bibfnamefont {A.}~\bibnamefont {Zaccone}},\ }\href
  {\doibase 10.1103/PhysRevB.100.220201} {\bibfield  {journal} {\bibinfo
  {journal} {Phys. Rev. B}\ }\textbf {\bibinfo {volume} {100}},\ \bibinfo
  {pages} {220201} (\bibinfo {year} {2019})}\BibitemShut {NoStop}%
\bibitem [{\citenamefont {Klinger}\ and\ \citenamefont
  {Kosevich}(2001)}]{Klinger}%
  \BibitemOpen
  \bibfield  {author} {\bibinfo {author} {\bibfnamefont {M.}~\bibnamefont
  {Klinger}}\ and\ \bibinfo {author} {\bibfnamefont {A.}~\bibnamefont
  {Kosevich}},\ }\href {\doibase https://doi.org/10.1016/S0375-9601(01)00090-1}
  {\bibfield  {journal} {\bibinfo  {journal} {Physics Letters A}\ }\textbf
  {\bibinfo {volume} {280}},\ \bibinfo {pages} {365} (\bibinfo {year}
  {2001})}\BibitemShut {NoStop}%
\bibitem [{\citenamefont {Goldstein}(1980)}]{goldstein}%
  \BibitemOpen
  \bibfield  {author} {\bibinfo {author} {\bibfnamefont {H.}~\bibnamefont
  {Goldstein}},\ }\href@noop {} {\emph {\bibinfo {title} {Classical
  mechanics}}},\ \bibinfo {edition} {2nd}\ ed.,\ Addison-Wesley series in
  physics\ (\bibinfo  {publisher} {Addison-Wesley},\ \bibinfo {address}
  {Reading, Mass. [etc},\ \bibinfo {year} {1980})\BibitemShut {NoStop}%
\bibitem [{\citenamefont {Brambilla}\ \emph {et~al.}(2020)\citenamefont
  {Brambilla}, \citenamefont {Leino}, \citenamefont {Petreczky},\ and\
  \citenamefont {Vairo}}]{Brambilla}%
  \BibitemOpen
  \bibfield  {author} {\bibinfo {author} {\bibfnamefont {N.}~\bibnamefont
  {Brambilla}}, \bibinfo {author} {\bibfnamefont {V.}~\bibnamefont {Leino}},
  \bibinfo {author} {\bibfnamefont {P.}~\bibnamefont {Petreczky}}, \ and\
  \bibinfo {author} {\bibfnamefont {A.}~\bibnamefont {Vairo}} (\bibinfo
  {collaboration} {TUMQCD Collaboration}),\ }\href {\doibase
  10.1103/PhysRevD.102.074503} {\bibfield  {journal} {\bibinfo  {journal}
  {Phys. Rev. D}\ }\textbf {\bibinfo {volume} {102}},\ \bibinfo {pages}
  {074503} (\bibinfo {year} {2020})}\BibitemShut {NoStop}%
\bibitem [{\citenamefont {Banerjee}\ \emph {et~al.}(2023)\citenamefont
  {Banerjee}, \citenamefont {Gavai}, \citenamefont {Datta},\ and\ \citenamefont
  {Majumdar}}]{Datta_2022}%
  \BibitemOpen
  \bibfield  {author} {\bibinfo {author} {\bibfnamefont {D.}~\bibnamefont
  {Banerjee}}, \bibinfo {author} {\bibfnamefont {R.}~\bibnamefont {Gavai}},
  \bibinfo {author} {\bibfnamefont {S.}~\bibnamefont {Datta}}, \ and\ \bibinfo
  {author} {\bibfnamefont {P.}~\bibnamefont {Majumdar}},\ }\href@noop {}
  {\enquote {\bibinfo {title} {Temperature dependence of the static quark
  diffusion coefficient},}\ } (\bibinfo {year} {2023}),\ \Eprint
  {http://arxiv.org/abs/2206.15471} {arXiv:2206.15471 [hep-ph]} \BibitemShut
  {NoStop}%
\bibitem [{\citenamefont {Hui}\ and\ \citenamefont {Skinner}(2023)}]{Skinner}%
  \BibitemOpen
  \bibfield  {author} {\bibinfo {author} {\bibfnamefont {A.}~\bibnamefont
  {Hui}}\ and\ \bibinfo {author} {\bibfnamefont {B.}~\bibnamefont {Skinner}},\
  }\href {\doibase 10.1103/PhysRevLett.130.256301} {\bibfield  {journal}
  {\bibinfo  {journal} {Phys. Rev. Lett.}\ }\textbf {\bibinfo {volume} {130}},\
  \bibinfo {pages} {256301} (\bibinfo {year} {2023})}\BibitemShut {NoStop}%
\end{thebibliography}%

\end{document}